\documentclass[onecolumn,groupedaddress,superscriptaddress,notitlepage]{revtex4-1}

\usepackage{amsfonts,amssymb,amsbsy,amsmath}
\usepackage{graphicx}
\usepackage{epstopdf}
\usepackage[colorlinks]{hyperref}
\hypersetup{pdftitle=Distress propagation in complex networks: the case of non-linear DebtRank, pdfauthor={Marco Bardoscia, Fabio Caccioli, Juan Ignacio Perotti, Gianna Vivaldo, Guido Caldarelli}}

\begin{document}

\title{Distress propagation in complex networks: the case of non-linear DebtRank}

\author{Marco Bardoscia}
\email[Email: ]{marco.bardoscia@gmail.com}
\affiliation{Department of Banking and Finance, University of Z\"{u}rich, Z\"{u}rich, Switzerland}
\affiliation{London Institute for Mathematical Sciences, London,  United Kingdom}

\author{Fabio Caccioli}
\email[Email: ]{f.caccioli@ucl.ac.uk}
\affiliation{University College London, Department of Computer Science, London, United Kingdom}
\affiliation{Systemic Risk Centre, London School of Economics and Political Sciences, London, United Kingdom}

\author{Juan Ignacio Perotti}
\email[Email: ]{juanpool@gmail.com}
\affiliation{IMT Institute for Advanced Studies, Lucca, Italy}

\author{Gianna Vivaldo}
\email[Email: ]{gianna.vivaldo@imtlucca.it}
\affiliation{IMT Institute for Advanced Studies, Lucca, Italy}

\author{Guido Caldarelli}
\email[Email: ]{guido.caldarelli@imtlucca.it}
\affiliation{IMT Institute for Advanced Studies, Lucca, Italy}
\affiliation{CNR-ISC: Institute of Complex Systems, Rome, Italy}
\affiliation{London Institute for Mathematical Sciences, London, United Kingdom}

%\date{\today}

\begin{abstract}
We consider a dynamical model of distress propagation on complex networks, which we apply to the study of financial contagion in networks of banks connected to each other by direct exposures. The model that we consider is an extension of the DebtRank algorithm, recently introduced in the literature. The mechanics of distress propagation is very simple: When a bank suffers a loss, distress propagates to its creditors, who in turn suffer losses, and so on. The original DebtRank assumes that losses are propagated linearly between connected banks. Here we relax this assumption and introduce a one-parameter family of non-linear propagation functions. As a case study, we apply this algorithm to a data-set of 183 European banks, and we study how the stability of the system depends on the non-linearity parameter under different stress-test scenarios. We find that the system is characterized by a transition between a regime where small shocks can be amplified and a regime where shocks do not propagate, and that the overall stability of the system increases between 2008 and 2013.
\end{abstract}

%\pacs{89.75.Hc,05.45.-a,89.75.-k,89.75.Fb}
%89.75.Hc Networks and genealogical trees
%05.45.-a Dynamical systems nonlinear, 
%89.75.-k Complex systems, 
%89.75.Fb Structures and organization in complex systems
%64.60.ah Percolation
%02.50.Ey: Stochastic processes 52,5
%05.45.Tp: Time series analysis
%05.40.-a Fluctuation phenomena, random processes, noise, and Brownian motion
%\keywords{Temporal Networks, Entropy, Burstiness}
\maketitle

%%%%%%%%%%%%%%%%%%%%  INTRODUCTION  %%%%%%%%%%%%%%%%%%%%%%%%%%%%%%%%%%%%%%%%%%%%%%%%%%%%%%%%%
\section{Introduction} \label{sec:intro}
Complex networks \cite{albert2002statistical,caldarelli2007scale-free,barabasi2011network} have proved useful to describe systems characterised by pair-wise interactions. Properties of dynamical processes on networks can be strongly affected by the underlying topology \cite{barrat2008dynamical}. Examples include spread of news \cite{leskovec2009meme}, rumours \cite{nekovee2007theory}, diseases \cite{pastor2014epidemic}, financial distress \cite{schweitzer2009economic}, random walkers travelling the graph \cite{motter2005network,ehrardt2006diffusion,scholtes2014causality}, and avalanches \cite{Watts2002asimple,lee2011impact}. 

In these cases stylized models, despite their apparent simplicity, can give meaningful indications on the large scale dynamics of the system \cite{pastor2014epidemic}, also helping to shed light on the importance of the  network topology \cite{vandenbroeck2011gleamviz}. For example, models of epidemic contagion (such as SIS or SIR \cite{allen1994some}) display dramatically different behaviors depending weather they take place on regular lattices or on complex networks. Similarly, also the spread of distress \cite{iori2006systemic,nier2007network,gai2010contagion,battiston2012liaisons} in financial networks is deeply dependent on the pattern of connections among financial institutions. In particular, it is not clear yet if a single topology can be considered robust with respect to different types of shocks \cite{tran2015mitigating} or not \cite{roukny2013default,caccioli2012heterogeneity}.

Financial institutions are strongly interconnected in a variety of ways (e.g.\ ownership relationships \cite{elliott2015financial,vitali2011thenetwork}, common asset holdings \cite{huang2013cascading,caccioli2014stability,guo2015thetopology}, trading of derivatives \cite{heise2012derivatives}, 
%\cite{heise2012derivatives,derrico2015hot}, 
possible arbitrage opportunities to exploit \cite{bardoscia2012financial}) through which distress can propagate and lead to amplification phenomena, such as default cascades. Here we focus on a single layer of interconnectedness, namely that associated with interbank loans. To cope with fluctuations of liquidity, banks constantly lend money to each other, at different maturities. Hence, lenders are subject to \emph{counterparty risk}, i.e.\ the risk that their borrowers could default and therefore not be able to fulfill their obligations. This, in turn, could lead to the default of lenders, resulting in a further wave of distress. 

\begin{figure}
\centering
\includegraphics[width=0.75\columnwidth]{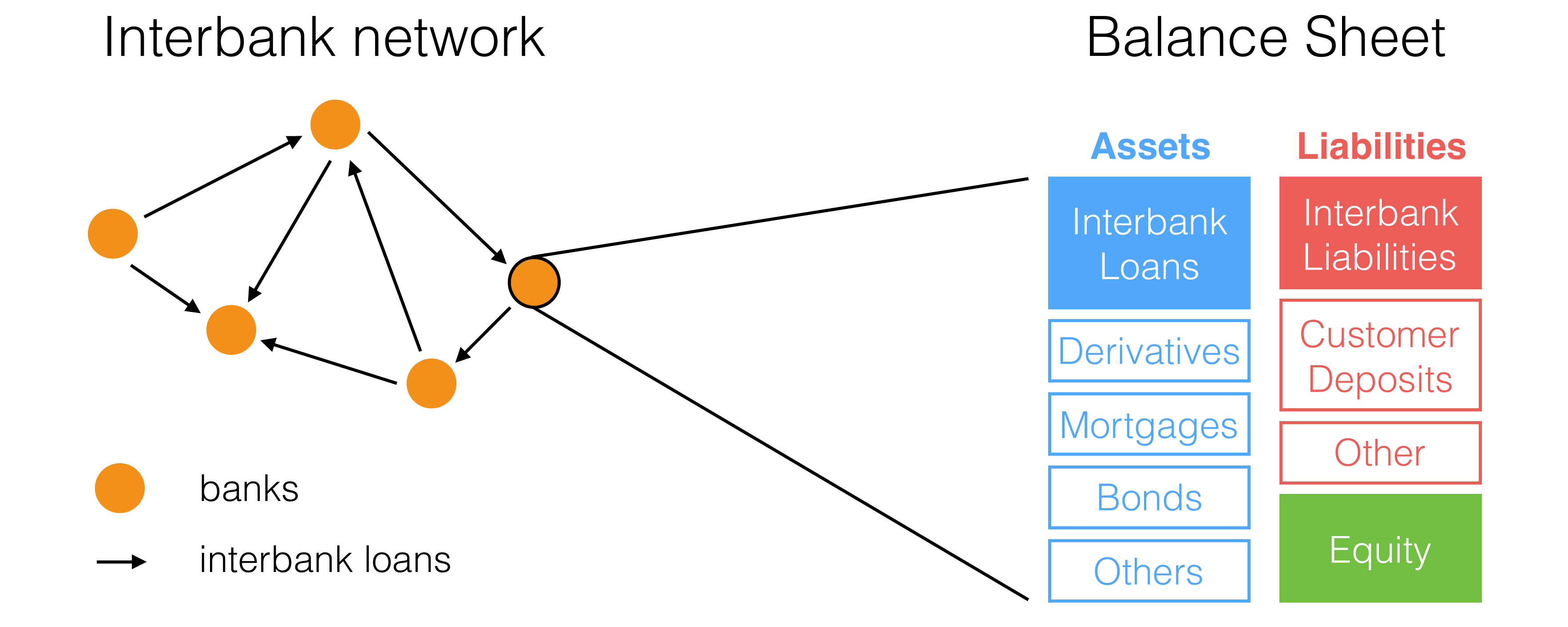}
\caption{\textbf{Interbank network and balance sheet.} Sketch of a portion of the interbank network and stylised representation of the balance sheet of a bank, with interbank assets and liabilities highlighted. The difference between assets and liabilities is the equity. A negative equity is usually considered a good proxy for the default of a bank.}
\label{fig:balancesheet}
\end{figure}

In the literature on financial contagion, a bank is represented by its balance sheet, consisting of \emph{assets} with a positive economic value (such as loans, derivatives, stocks, bonds, real estate) and of \emph{liabilities} with a negative economic value (such as customers' deposits, debits). The \emph{balance sheet identity} for bank $i$ defines its \emph{equity} as the difference between its total assets $A_i$ and its total liabilities $L_i$: $E_{i}=A_{i}-L_{i}$. A bank with a negative equity would not be able to pay back its debtors, even assuming that it could sell all of its assets. Therefore, usually a negative equity is considered a good proxy for the default of a bank. An interbank loan extended by bank $i$ to bank $j$ is an asset for bank $i$ and a liability for bank $j$. Hence, the relationship between a lender and a borrower is pair-wise in nature and a convenient way to represent it is by means of a directed weighted network \cite{demasi2006fitness,cont2010network} in which edges of weight $A_{ij}^{IB}$ correspond to a loan of amount $A_{ij}^{IB}$ from bank $i$ to bank $j$. We call all the other assets and liabilities \emph{external} and we denote them with $A_i^E$ and $L_i^E$ respectively (see Fig.\ \ref{fig:balancesheet}).

The study of the interbank network has attracted considerable attention, also for its practical importance. Two widely recognized algorithms to quantify losses due to financial contagion are the Furfine algorithm \cite{Watts2002asimple,furfine2003interbank} and DebtRank \cite{battiston2012debtrank,battiston2013capital,thurner2013debtrank,battiston2015leveraging,battiston2015debtrank,battiston2015nexus,bardoscia2015debtrank}. The former is essentially a threshold model according to which a bank propagates distress to its creditors only after its default. In contrast, DebtRank was introduced precisely to account for shock propagations occurring also in absence of default. To this end, relative losses in the equity of a borrower translate into the same relative devaluation of interbank assets of the corresponding lender. Those two mechanisms represent two extremes. On one hand, the Furfine algorithm is likely to underestimate the build-up of systemic risk. On the other hand, in DebtRank even tiny variations in the equity (as those deriving from daily market fluctuations) have a sizeable impact on the value of interbank assets. As a realistic scenario is likely to lie in-between those two extremes, in this paper we propose a model that interpolates between them and use it to perform stress tests to the European banking system. We will refer to the introduced model as non-linear DebtRank.

The paper is organized as follows: In Section \ref{sec:results} we specify the model used and present a detailed characterization of its behavior within the context of a case study. In Section \ref{sec:discussion} we discuss the main implications of our results, also from a policy-making perspective, and point out some limitations of our approach. We refer the reader interested in the details about the data used and a derivation of the algorithm to Section \ref{sec:methods}.

% FIGURE 1
% END FIGURE 1

%%%%%%%%%%%%%%%%%%  SEC: RESULTS  %%%%%%%%%%%%%%%%%%%%%%%%%%%%%%%%%%%%%%%%%%%%%%%%%%%%%%%%%%%%%%%%%
\section{Results} \label{sec:results}
We perform stress tests on $N=183$ publicly traded European banks using data from their balance sheets for the years from $2008$ to $2013$ (see 
%Section \ref{sec:methods} 
the section Materials and Methods: Data 
for a detailed description of data). Since data on bilateral exposures are not publicly available, we employ a reconstruction technique to infer plausible values \cite{Cimini2014a,Cimini2014b} and sample for each year 100 instances of interbank networks for given values of connectivity $p$, defined as the number of reconstructed edges divided by the number of possible edges ($N(N-1)$) (see 
%Section \ref{sec:methods} 
the section Materials and Methods: Data 
for more details about the reconstruction of data). 

A stress test consists in applying an initial exogenous shock to the system and to measure its response in terms of the resulting equity losses.
From the point of view of risk management, the relevant quantity is the \emph{relative equity loss} of bank $i$ at time $t$:
\begin{equation} \label{eq:h}
h_i(t) = \frac{E_i(0) - E_i(t)}{E_i(0)} \, . 
\end{equation}
The corresponding quantity at the aggregate level is the \emph{total relative equity loss}:
\begin{equation}
H(t) = \frac{\sum_{i=1}^N E_i(0) - E_i(t)}{\sum_{i=1}^N E_i(0)} 
= \sum_{i=1}^N \left( h_{i}(t)\frac{E_{i}(0)}{\sum_{j=1}^N E_{j}(0)}\right) \, .
\end{equation}
In the context of financial contagion, the initial exogenous shock amounts to a relative devaluation of external assets, which corresponds to setting the initial condition $h_i(1)$ (see the section Materials and Methods: Model dynamics). If banks hold no interbank asset they will not incur any additional equity loss and their final equity losses will be equal only to the devaluation of external assets. In contrast, let us suppose that bank $i$ extends a loan to bank $j$ and therefore that it holds the corresponding interbank asset $A_{ij}^{IB}$. If bank $j$ suffers an equity loss, its probability of default will increase and thus the probability that it will be able to fully pay back its loan will decrease. Bank $i$ will account for the possibility of not being fully paid back by reducing the value of the interbank asset $A_{ij}^{IB}$ in its balance sheet (in financial jargon the interbank asset will be marked-to-market). However, as the value of the interbank asset $A_{ij}^{IB}$ decreases, also the equity of bank $i$ will decrease by the same amount. Lenders of bank $i$ will now re-evaluate their interbank assets towards bank $i$, using the same mechanism used by bank $i$ to re-evaluate its interbank asset towards bank $j$. As a consequence, the process of re-evaluation of interbank assets and equities proceeds recursively from borrowers to lenders, until convergence.

\begin{figure*}
\centering
\includegraphics[width=0.36\linewidth]{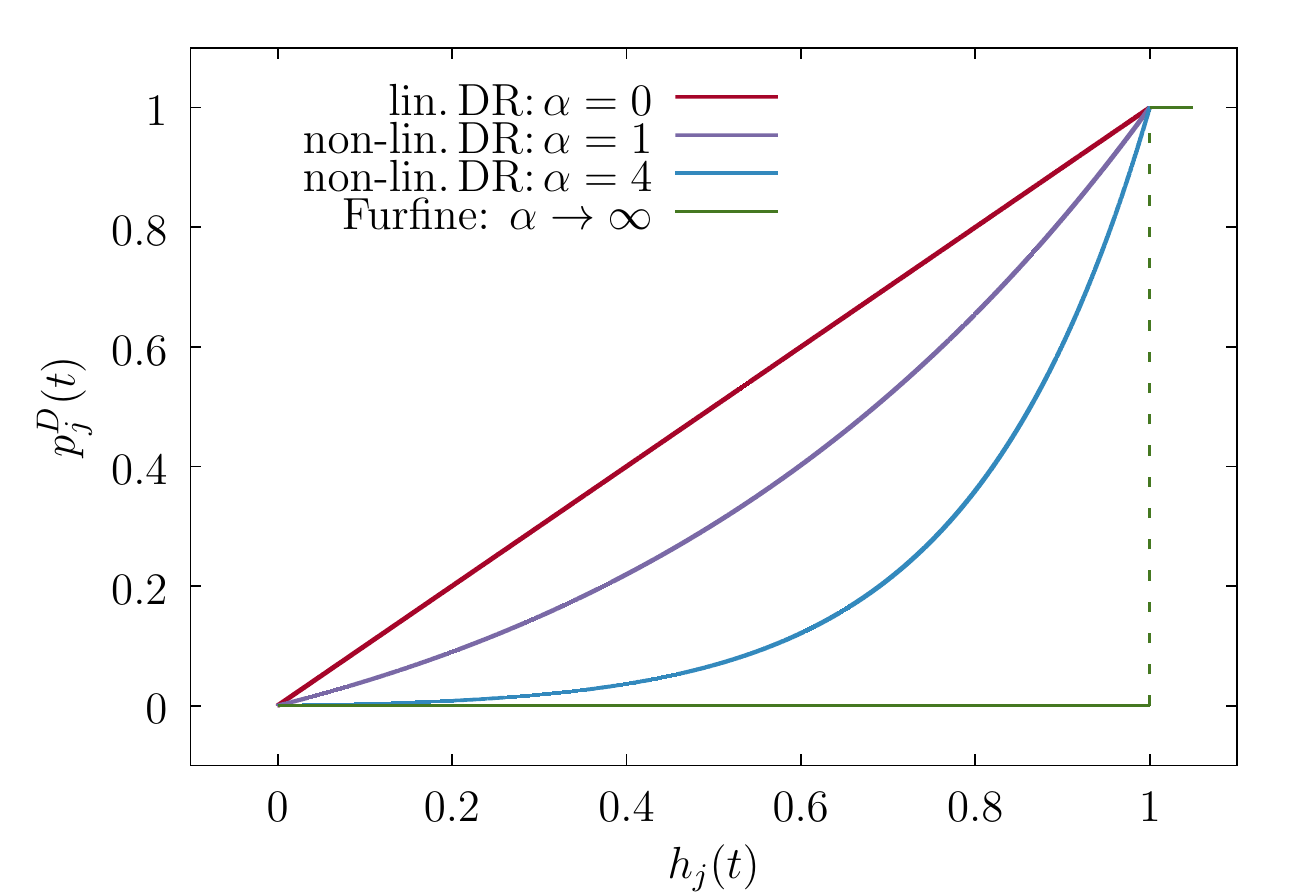}
\caption{\textbf{Comparison of different algorithms.} Probability of default $p_j^D(t)$ as a function of the relative equity loss $h_j(t)$ for different algorithms. The non-linear DebtRank interpolates between the Furfine algorithm and the non-linear DebtRank.}
\label{fig:pd_vs_h} 
\end{figure*}

As we show in 
%Section\ \ref{sec:methods} 
the section Materials and Methods: Model dynamics,  
the dynamic equation for the relative equity loss that includes the aforementioned recursive re-evaluation of interbank assets is the following:
\begin{equation} \label{eq:map}
h_i(t+1) = \min \left\{ 1, h_i(t)\!+\!\sum_j \Lambda_{ij} \left[ p^\mathrm{D}_j(t) - p^\mathrm{D}_j(t-1)\right] \right\} \, ,
\end{equation}
where $p^\mathrm{D}(t)$ is the probability of default of bank $j$ at time $t$ and $\Lambda_{ij} = A_{ij}^{IB} / E_i(0)$ is the interbank leverage matrix. 
In order to completely define the iterative map \eqref{eq:map}, we need to establish a relationship between $p^\mathrm{D}_j(t)$ and $h_j(t)$. In the case of Furfine's algorithm, the probability of default is equal to one only if the equity is smaller than or equal to zero, and it is equal to zero otherwise, while in the linear DebtRank $p^\mathrm{D}_j(t) = h_j(t)$. Here we take a practical approach by introducing the following functional form:
\begin{equation} \label{eq:p_i}
p^\mathrm{D}_{j}(t) = h_{j}(t)e^{\alpha \left[h_{j}(t)-1\right]} \, ,
\end{equation}
in which $\alpha$ is a free parameter. Such approach, albeit ``phenomenological'' in the sense that \eqref{eq:p_i} is intended to provide an effective description of how shocks propagate, has several merits. It depends on a single parameter, which has a clear financial meaning: it is the inverse of the typical relative equity loss after which banks start to propagate distress to their creditors. Thus, $1/\alpha$ can be interpreted as a \emph{soft threshold}: losses smaller that $1/\alpha$ have negligible impact on the probability of the default of banks. The value $\alpha$ should be therefore tuned by practitioners and regulators so that the probabilities of default can be calibrated (in principle also diversely) for each bank. Estimating it is beyond the scope of this paper. However, we will perform a sensitivity analysis by exploring the behaviour of the model as a function of $\alpha$. Moreover, \eqref{eq:p_i} leads to probabilities of defaults that are convex with respect to the equity (see Fig.\ \ref{fig:pd_vs_h}), which is their expected behaviour. In fact small fluctuations in the equity of borrowers reasonably trigger tiny variations in the probability of the default of lenders, while when borrowers experience larger losses the marginal effect on lenders can be dramatic. Finally, \eqref{eq:p_i} easily allows to interpolate between two of the mostly widely used contagion models: the linear DebtRank, which is recovered for $\alpha = 0$, and the Furfine algorithm, which is recovered for $\alpha \to \infty$. In Fig.\ \ref{fig:pd_vs_h} we plot for these cases the probabilities of default as a function of the relative equity loss. The importance of the interbank leverage matrix in the context of distress propagation has been already highlighted by \cite{Markose2012,battiston2015leveraging,bardoscia2015debtrank}, pointing out that the stability of the system is determined by its largest eigenvalue. Also in our case, using the general framework discussed in \cite{bardoscia2016pathways} it is easy to show that the system is stable if $\lambda_{\text{max}} e^{\alpha} < 1$, i.e.\ if $\alpha > \log{\lambda_{\text{max}}}$, where $\lambda_{\text{max}}$ is the largest eigenvalue of the interbank leverage matrix $\Lambda$.

Our stress test consists in assuming a devaluation of external assets by a factor $x_{\text{shock}}$ for a fraction of banks equal to $p_{\text{shock}}$. All presented results are averaged both over the network samples and over the set of initially shocked banks (10 realizations of such set for each network in the sample). Therefore, independently for each realization, we proceed to shock $p_{\text{shock}} \cdot N$ banks. Each shocked bank suffers an initial loss in the external assets: $A_i^E(1) = x_{\text{shock}} A_i^E(0)$ (see 
%Section \ref{sec:methods} for more details).
the section Materials and Methods: Model dynamics). 

\begin{figure*}
\centering
\includegraphics*[width=\columnwidth]{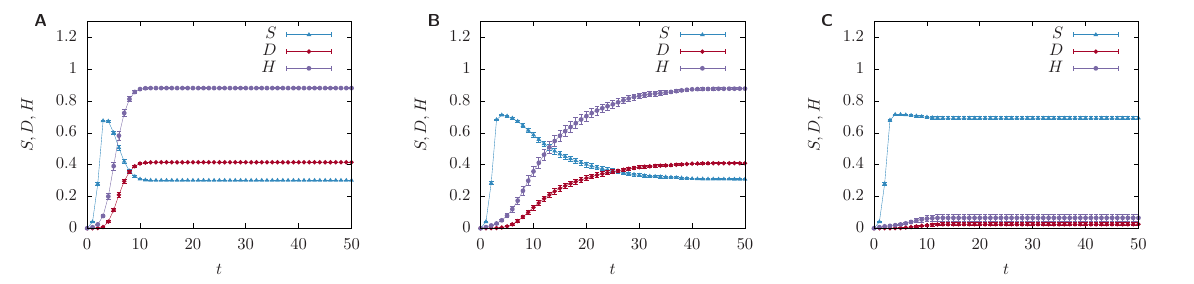}
\caption{\textbf{Unraveling of shocks propagation over time.} Fraction of stressed banks $S(t)$ (blue line), fraction of defaulted banks $D(t)$ (red line), and $H(t)$ (violet line), total relative equity loss experienced by the system as a function of the time $t$ over which shocks propagate. Banks experience a shock in the external assets, which suffer a relative loss equal to $x_{\mathrm{shock}}=0.5\%$. All points are averaged over a sample of 100 reconstructed networks with connectivity $p=0.05$ and compatible with 2008 balance sheets, and over 10 realisations of the shock in which each bank is shocked with probability $p_{\mathrm{shock}}=0.05$. Error bars span three standard errors. $\alpha = 0$ in panel A and the algorithm reduces to the linear DebtRank, while $\alpha = 1$ in panel B, and $\alpha=2$ in panel C. We see that the dynamics unravels within a few time steps in the panels A and C, while it takes considerably more time steps in panel B.}
\label{fig:2008HASD} 
\end{figure*}

A first analysis is focused on the most critical year: $2008$. In Fig.\ \ref{fig:2008HASD} we show the effects of the initial shock, as its propagation through the network unravels in time. More in details, Fig.\ \ref{fig:2008HASD} shows $S(t)$, the fraction of stressed banks, $D(t)$, the fraction of defaulted banks, and $H(t)$, the total relative equity loss experienced by the system. Stressed banks are those which have experienced equity losses, but have not defaulted yet. Defaulted banks at time $t$ are those for which $h_i(t) = 1$. From left to right, plots correspond to $\alpha=0$, $\alpha=1$, and $\alpha=2$.

% FIGURE 2 % % % %
% END OF FIGURE 2

The qualitative behaviour of both stressed and defaulted banks is shared by all panels. Stressed banks sharply increase in the first time steps and decrease afterwards, as defaults start to occur. This is consistent with the fact that stress propagates even in absence of defaults. However, a clear dependence from $\alpha$ emerges. The most striking feature is that the time scale over which the system reaches its steady state is a non-monotonous function of $\alpha$. In fact, in both panels A and C convergence is reached before the first 20 time steps, while in panel B the dynamics is slower. 
This phenomenon can be intuitively understood as follows: $\alpha$ is related to a soft threshold in the value of the relative equity loss that a bank needs to attain before it can propagate a shock. Such threshold is zero in the linear case and approaches one in the strongly non-linear regime ($\alpha \gg 1$). In the first case shocks are quickly propagated through the network, while in the second case shocks are easily dampened. In the intermediate regime the build-up of the stress happens gradually. Nevertheless, the total relative equity loss can still reach values comparable to those of the linear case (see panels A and B).

\begin{figure*}
\centering
\includegraphics*[width=\columnwidth]{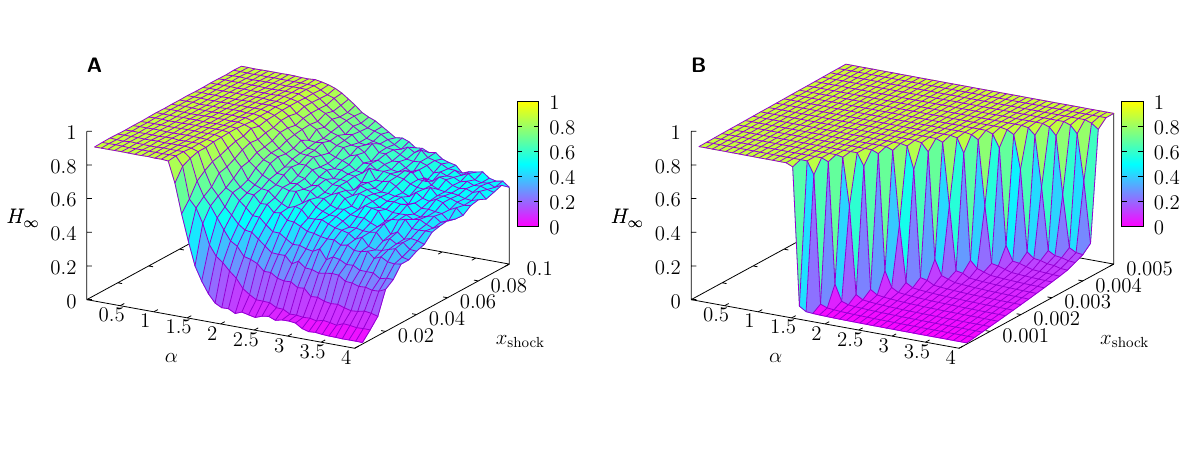}
\caption{\textbf{Equity loss in 2008.} Surface plot of $H_\infty$, the total relative equity loss in the steady state, as a function of the size of the shock suffered by external assets of banks $x_{\text{shock}}$ and of the parameter $\alpha$, which tunes the non-linearity of the algorithm. All points are averaged over a sample of 100 reconstructed networks with connectivity $p=0.05$, and compatible with 2008 balance sheets, and over 10 realization of the shock in which each bank is shocked with probability $p_{\mathrm{shock}}=0.05$ (panel A) and $p_{\mathrm{shock}}=1$ (panel B). Note that the range of the total size of the shock $p_{\mathrm{shock}} \cdot x_{\mathrm{shock}}$ is the same for both panels. As $\alpha$ increases, the propagation of the shock is dampened, resulting in smaller losses, and in two different regimes, whose separation is especially evident for $p_{\mathrm{shock}}=1$, i.e.\ when all banks are shocked.}
\label{fig:H3D_2008} 
\end{figure*}

Next, we present a comprehensive characterization of the steady state. Fig.\ \ref{fig:H3D_2008} shows the surface plots of $H_\infty$, the total relative equity loss in the steady state, as a function of $\alpha$ and $x_\mathrm{shock}$, for network connectivity $p=0.05$, and for different choices of $p_\mathrm{shock}$. n order to meaningfully compare the results for different values of $p_\mathrm{shock}$, we tune the range of $x_\mathrm{shock}$ spanned so that the ranges of total shock $p_{\mathrm{shock}} \cdot x_{\mathrm{shock}}$ affecting the system are equal across all the cases considered.

% FIGURE 3 % % % %
% END FIGURE 3

Let us start from focusing on the behaviour of $H_\infty$ as a function of $\alpha$ for fixed values of $x_\mathrm{shock}$. From the panel A of Fig.\ \ref{fig:H3D_2008} we can clearly see that, apart from the numerical fluctuations due to the finite number of realizations, for any given value of $x_\mathrm{shock}$, $H_\infty$ decreases monotonically with $\alpha$. Starting from large values of $\alpha$ and moving towards smaller values, $H_\infty$ reaches a plateaux in which most of the equity in the system has been lost ($H_\infty > 0.9$). As expected, focusing instead on the behaviour of $H_\infty$ as a function of $x_\mathrm{shock}$, for fixed values of $\alpha$, $H_\infty$ increases with $x_\mathrm{shock}$. Overall we can detect two different regimes, one in which (almost) all the equity in the system has been lost (for small values of $\alpha$) and one in which (almost) no equity has been lost (for large values of $\alpha$ and small values of $x_\mathrm{shock}$). For smaller and smaller values of $x_\mathrm{shock}$ the crossover between such two regimes is sharper and sharper. Finally, in the limit $x_\mathrm{shock} \to 0$ the system displays a transition between a stable regime in which no losses occur, and an unstable regime in which also infinitesimal shocks can lead to large total relative losses. The presence of such transition can be better appreciated from panel B of Fig.\ \ref{fig:H3D_2008}, where we present the results in the case in which all banks suffer the same initial shock (i.e.\ $p_\mathrm{shock} = 1$). This can be interpreted as a shock to a risk factor common to all banks, such as a sudden change in interest rates or similar to that experienced during a major macroeconomic downturn. We note here that we have performed simulations for different values of the connectivity parameter $p$ ranging between $p=0.05$ to $p=1$ (fully connected network). Interestingly, we observed that systems with very different connectivities behave in a similar way. A possible explanation is that, due to the reconstruction technique used (see 
%Section \ref{sec:methods}), 
the section Materials and Methods: Data), 
$p = 0.05$ is already enough to connect systemically important banks.

\begin{figure*}
\centering
\includegraphics*[width=\columnwidth]{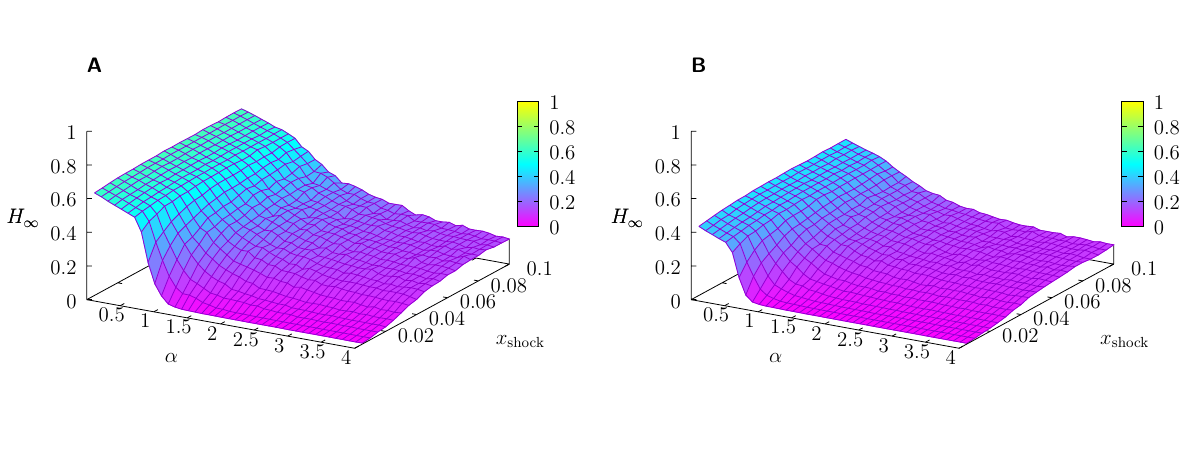}
\caption{\textbf{Equity loss in 2010 and 2013.} Analogous of Fig.\ \ref{fig:H3D_2008}, but for the years 2010 and 2013. Here the connectivity of the reconstructed networks is $p=0.05$ and the fraction of shocked banks on average is $p_{\mathrm{shock}}=0.05$. Going from $2008$ (see Fig.\ \ref{fig:H3D_2008}, panel A) to $2013$ $H_\infty$ is less and less sensitive to changes in $\alpha$ and it is generally smaller, meaning that banks are more and more robust.}
\label{fig:H3D_years}
\end{figure*}

Finally, in Fig.\ \ref{fig:H3D_years} we adopt the same setting as in Fig.\ \ref{fig:H3D_2008} for different years. Overall, we observe that $H_\infty$ markedly decreases from 2008 to 2013. It clearly emerges that the system was more prone to amplify shocks in $2008$, when a region in parameters space in which $H_\infty \simeq 1$ exists. This is consistent with the intuition that banks in $2008$ were more fragile.

% FIGURE 4 (H3D2008, H3D2010, H3D2013)
% END FIGURE 4

%%%%%%%%%%%%%%%%%%  SEC: DISCUSSION  %%%%%%%%%%%%%%%%%%%%%%%%%%%%%%%%%%%%%%%%%%%%%%%%%%%%%%%%%%%%%%%%%
\section{Discussion} \label{sec:discussion}
In the present study, a general shock propagation mechanism is applied to an interbank network of 183 publicly traded European banks. With probability $p_\mathrm{shock}$ each bank is subject to an initial shock consisting in the devaluation of their external assets by a factor $x_\mathrm{shock}$. The system reaction to shocks is measured in terms of the total relative equity loss, which takes into account the contribution of each bank to the relative equity loss of the system. The dynamics of shock propagation that we consider (non-linear DebtRank) interpolates through the parameter $\alpha$ between two stress test algorithms: the Furfine algorithm and the linear DebtRank.

We notice that the propagation of shocks strongly depends on the parameter $\alpha$. In particular, in all stress scenarios that we have considered we observe a crossover between a regime of large losses (for small $\alpha$), in which potentially all banks could default, and a regime of small losses (for large $\alpha$), in which most banks survive the shock. The width of the intermediate region shrinks as the fraction of banks affected by the shock approaches one.

The model also shows that the interbank network was significantly more fragile in 2008, when the financial crisis took place, than in the subsequent years.
This observation holds qualitatively for all values of model parameters and  connectivities explored and it is also in agreement with other empirical studies \cite{memmel2013contagion}.

In addition to the properties of the steady state we have also looked into the dynamics of quantities such as the number of stressed and defaulted banks, whose behavior highlights the existence of different time scales, depending on the model parameters. For instance, we observe that the time needed to reach the steady state is a non-monotonic function of $\alpha$: in certain cases the shock produces a slow drive of the interbank network towards its collapse, while in other cases the crash occurs immediately after the shock. 

Clearly, establishing a coherent mapping between probability of defaults and changes in equity opens several possible directions for future research. Obviously, calibrating $\alpha$, possibly extracting a different value for each bank, would represent a major achievement. Beyond the ``phenomenological'' approach adopted here, one could try to derive such relationship in the context of standard financial risk management theory. Moreover, here we have limited our analysis to direct exposures due interbank landing. A proper assessment of systemic risk should account for additional types of interconnectedness, such as that associated with overlapping portfolios, exchange of derivates, and ownership structure. Hence, another future extension of the model could be based on a multilayer network that incorporates those effects. Complex interactions across different layers could lead to non-trivial amplification phenomena \cite{caccioli2015overlapping}. Finally, studying the stability of the system, which is becoming increasingly possible due to recent works \cite{bardoscia2016pathways,tan2016symmetric}, could allow regulators to develop faster qualitative stress-testing frameworks to continuously monitor systemic risk.

%%%%%%%%%%%%%%%%%%  SEC: METHODS  %%%%%%%%%%%%%%%%%%%%%%%%%%%%%%%%%%%%%%%%%%%%%%%%%%%%%%%%%%%%%%%%%
\section{Materials and Methods}\label{sec:methods}

\subsection{Model dynamics}
In order to derive \eqref{eq:map} we start from the balance sheet identity introduced in 
%Section \ref{sec:intro} 
the section Introduction 
in which we distinguish between external and interbank assets and liabilities:
\begin{equation} \label{eq:equity_t}
E_i(t) = A_i^E(t) - L_i^E + \sum_{j=1}^N A_{ij}^{IB}(t) - L_{ij}^{IB} \, ,
\end{equation}
where, as customary in the literature on financial contagion, assets depend explicitly on time, while liabilities do not. In fact, while external assets are subject to exogenous fluctuations, the value of interbank assets $A_{ij}^{IB}$ will change accordingly (marked-to-market), depending on the probability of default of the borrower $j$.
However, the fact that bank $i$ reassesses the value of its interbank claim towards bank $j$ does not change the value of the debt that bank $j$ owes to bank $i$: hence interbank liabilities (and analogously external liabilities) always keep their face value (i.e.\ the value initially present in the balance sheet), and therefore do not depend on time. As a consequence, using \eqref{eq:h} and \eqref{eq:equity_t} we can readily compute:
\begin{equation} \label{eq:h_1}
h_i(t) = \frac{A_i^E(0) - A_i^E(t)}{E_i(0)} + \sum_{j=1}^N \frac{A_{ij}^{IB}(0) - A_{ij}^{IB}(t)}{E_i(0)} \, .
\end{equation} 
If the probability of default of bank $j$ is equal to zero, obviously its creditor bank $i$ can expect to fully recover the face value $A_{ij}^{IB}(0)$ of its loan. In contrast, if the probability of default of bank $j$ is equal to one, bank $j$ has defaulted and its creditor bank $i$ can expect to recover only a fraction $R_{ij} A_{ij}^{IB}(0)$ of its loan, where $R_{ij} \in [0, 1]$. A natural way to interpolate between these two extreme cases is to identify the value of interbank assets at time $t + 1$ with their expected value at time $t$:
\begin{equation}
A_{ij}^{IB}(t+1) = A_{ij}^{IB}(0) \left( 1 - p_j^D(t) \right) + R_{ij} A_{ij}^{IB}(0) p_j^D(t) \, .
\end{equation}
By making the conservative assumption \cite{gai2010contagion,markose2012systemic} that creditors do not actually recover any amount of money in case of default of their borrowers, we have: 
\begin{equation} \label{eq:ib_assets}
A_{ij}^{IB}(t+1) = A_{ij}^{IB}(0) \left( 1 - p_j^D(t) \right) \, ,
\end{equation}
which implies that bank $i$ updates the value of its interbank claims towards bank $j$ such that it is equal to their face value if the probability of default $p_j^D$ of the borrower bank $j$ is zero and it decreases proportionally to $p_j^D$ otherwise. By plugging \eqref{eq:ib_assets} into \eqref{eq:h_1} and using the definition of the interbank leverage matrix \cite{battiston2015leveraging,bardoscia2015debtrank}:
\begin{equation}
\Lambda_{ij} = \frac{A_{ij}^{IB}(0)}{E_i(0)} \,
\end{equation}
we can immediately compute:
\begin{equation} \label{eq:delta_h}
h_i(t+1) - h_i(t) = \frac{A_i^E(t) - A_i^E(t+1)}{E_i(0)} + \sum_{j=1}^N \Lambda_{ij} \left[ p_j^D(t) -  p_j^D(t-1) \right] \, .
\end{equation} 
In our stress test scenario we will initially shock external assets by a relative amount $x_{\text{shock}}$, i.e.\ $A_i^E(1) = x_{\text{shock}} A_i^E(0)$, such that $h_i(1) = (1-x_{\text{shock}}) \left( A_i^E(0) / E_i(0) \right)$. However, after the initial shock external assets do not change anymore and the evolution of the relative equity losses is entirely due to the re-assessment of the value interbank assets. As a consequence, the first term in the right-hand side of \eqref{eq:delta_h} is equal to zero, for $t > 1$. Finally, when the equity of bank $i$ becomes equal to zero the bank defaults and is not able neither to further propagate shocks nor to sustain any additional losses. Hence, the maximum value attainable by the relative equity losses is one, which leads to \eqref{eq:map}.

From \eqref{eq:map}, we see that the results will depend on the relationship that we assume between the relative loss in equity of a bank and its probability of default. In the Furfine algorithm we have that:
\begin{equation}
p_j^D(t) = 
\begin{cases}
0 & \text{if \,} h_j(t) < 1 \\
1 & \text{if \,} h_j(t) = 1 \, ,
\end{cases}
\end{equation}
while in the linear DebtRank:
\begin{equation}
p_j^D(t) = h_j(t) \, .
\end{equation}
In the non-linear DebtRank we interpolate between this two extreme cases by means of the parameter $\alpha$ (see \eqref{eq:p_i}): for $\alpha = 0$ we recover the linear DebtRank, while for $\alpha \to \infty$ we recover the Furfine algorithm.

\subsection{Data}
The original source for raw data about balance sheets of banks is the Bureau van Dijk Bankscope database, from which we extract data for a subset of 183 among the largest European banks. In particular we use data about the total interbank assets (liabilities) $A_i^{IB} = \sum_{j=1}A_{ij}^{IB}$ ($L_i^{IB} = \sum_{j=1}L_{ij}^{IB}$) and we compute external assets (liabilities) as the difference between total assets (liabilities) and interbank assets (liabilities): $A_i^E = A_i - A_i^{IB}$ ($L_i^E = L_i - L_i^{IB}$). The same data have already been used in \cite{battiston2015leveraging,bardoscia2015debtrank}. See \cite{battiston2015leveraging} for all the details about the handling of missing data. Even though European banks do not necessarily constitute an isolated system \cite{kubelec2010geographical}, they are weakly correlated to American and Asian banks \cite{sandoval2014structure}. Moreover, banks outside Europe fall under different  regulating authorities, and therefore they might be subject to different conditions (e.g.\ during bailouts). The analysis of Ref.\ \cite{sandoval2014structure} makes it clear that European banks are also connected to financial institutions other than banks. To account for the additional channels of contagion due to the interaction between financial institutions of different types, one could extend our approach to the case of multilayer networks, as suggested in the section Discussion.

As already pointed out, the balance sheet of bank $i$ contains only data about its total interbank lending and borrowing, i.e.\ the values of $A_i^{IB}$ and $L_i^{IB}$. As a consequence, the full matrix needs to be reconstructed by making some assumptions. Here we proceed as in \cite{bardoscia2015debtrank} and make the reasonable assumption, successfully tested in \cite{Cimini2014a,Cimini2014b}, that the probability that bank $i$ extended a loan to bank $j$ is proportional to $A_i^{IB}$, the total amount of interbank lending of bank $i$ and to $L_i^{IB}$, the total amount of interbank borrowing of bank $j$. The fitness model \cite{musmeci2013bootstrapping} allows us to compute the values of the probabilities $\{p_{ij}\}$ that an edge $i \rightarrow j$ exists for a given value of connectivity $p$. We then use $\{p_{ij}\}$ to build a sample of 100 directed un-weighted networks for each year. For each network in the sample we assign weights using the iterative RAS algorithm \cite{upper2004estimating}. See \cite{bardoscia2015debtrank} for a full account of the procedure.

%%%%%%%%%%%%%%%%%%  ACKNOWLEDGMENTS  %%%%%%%%%%%%%%%%%%%%%%%%%%%%%%%%%%%%%%%%%%%%%%%%%%%%%%%%%%%%%%%%%
\section*{Acknowledgments}
All authors thank Marco D'Errico for sharing the data and for fruitful discussions and Stefano Battiston for the numerous exchanges of views. MB, JIP, GV, and GC acknowledge support from FP7-ICT project MULTIPLEX nr.\ 317532, FP7-ICT project SIMPOL nr.\ 610704, and Horizon 2020 project DOLFINS nr.\ 640772. FC acknowledges support of the Economic and Social Research Council (ESRC) in funding the Systemic Risk Centre (ES/K002309/1). GC acknowledges additional support from Horizon 2020 projects SoBIGData nr.\ 654024 and CoEGSS nr.\ 676547.

\section*{Author contributions statement}
Contributed to designing the model and the numerical experiments: MB, FC, JIP, GV, and GC. Run the numerical simulations: JIP and GV. Discussed the results and written the paper: MB, FC, JIP, GV, and GC.

\section*{Additional information}
\textbf{Competing financial interests.} The authors declare no competing financial interests. \textbf{Source code.} All the analysis has been developed in Python and is available upon request to the authors.

%%%%%%%%%%%%%%%%%%  REFERENCES  %%%%%%%%%%%%%%%%%%%%%%%%%%%%%%%%%%%%%%%%%%%%%%%%%%%%%%%%%%%%%%%%%
%\bibliographystyle{apsrev4-1}
\bibliographystyle{plos2015}
\bibliography{Debtrank.bib}

\end{document}